\documentstyle[12pt,newpasp,twoside,epsf]{article}
\def\edcomment#1{\iffalse\marginpar{\raggedright\sl#1\/}\else\relax\fi}
\reversemarginpar

\begin{document}
\title{Ionization and Velocity Structure in the Supernova Remnant~E0102-72}
 \author{K.A.~Flanagan, C.R.~Canizares, D.S.~Davis, D.~Dewey, J.C.~Houck,
T.H.~Markert, M.L.~Schattenburg}
\affil{Center for Space Research, Massachusetts Institute of Technology, \\
77 Massachusetts Ave., Cambridge, MA 02139}

\begin{abstract}The High Energy Transmission Grating (HETG) 
Spectrometer aboard the
Chandra X-Ray Observatory was used to observe E0102-72, a $\sim$1000 year old,
oxygen rich supernova in the Small Magellanic Cloud. The HETG disperses
the image of the remnant into a spectrum of images in the light of
individual X-ray emission lines.  Doppler shifts in the strongest lines
of oxygen and neon reveal bulk motions of up to 2000 km/sec with a 
complex morphology. Comparison of progressive ionization stages of magnesium,
neon, oxygen and silicon provide new insights into the mechanism of 
the `reverse shock' that heats the stellar ejecta.
\end{abstract}

\section{Introduction}

1E0102.2-7219 is a young ($\sim$1000 years) supernova remnant (SNR) 
in the Small Magellanic Cloud. It was discovered in X-rays 
with the {\it Einstein}
Observatory (Seward \& Mitchell 1981), and shortly afterwards optical 
filaments of oxygen were found (Dopita et al. 1981) and measured 
to have velocities of thousands of km/s (Tuohy \& Dopita 1983), pegging it as 
one of a small number of identified oxygen-rich SNRs.
As a class, these SNRs are believed to come from supernovae with 
massive progenitors, exhibit high velocities in their optical filaments, 
are young (with the possible exception of Puppis A) and often show 
evidence of asymmetrical explosion. E0102-72 has been observed in UV 
(Blair et al. 1989; Blair et al. 2000) and radio 
(Amy \& Ball 1993), and in the X-ray with {\it Chandra} and
other X-ray observatories (Hayashi et al. 1994; 
Gaetz et al. 2000; Hughes et al. 2000). Based on early X-ray
observations, Hughes (1988) ruled out a uniform-density shell model. 
Our data (Houck et al. 2000) and Gaetz et al. (2000)
confirm Hughes' conclusion. Gaetz et al. (2000) have looked at direct 
{\it Chandra} images of E0102-72 and noted the radial variation with 
energy bands centered on OVII and OVIII, suggesting an ionizing shock 
propagating inward. The analysis of the high resolution dispersed spectrum, 
discussed below, reinforces this interpretation.
In addition, Doppler shifts in the high-resolution spectrum give us
a first look at the velocity structure.


\section{The Observations}

	The supernova remnant E0102-72 was observed with the {\it Chandra}
High Energy Transmission Gratings (HETG) in two segments as part 
of the guaranteed time observation program. 
The configuration included HETG with the Advanced CCD Imaging Spectrometer 
(ACIS-S). ACIS has moderate spectral resolution which allows order sorting
of the grating spectrum.  The two E0102-72 observations had slightly 
different roll angles and aim points. Most of the analysis below was 
therefore restricted to a single observation (Obsid 120).

\begin{table}[h]
\large{\caption{HETG observations of E0102-72.}}
\label{tab:obs}
\begin{center}
\begin{tabular}{|c|c|} \hline\hline

\large{{Obsid 120}} & \large{{Obsid 968}}   \\ \hline
\large{{90 ksec}} & \large{{48.6 ksec}}  \\ \hline
\large{{Sept 28, 1999}} &\large{{Oct 8, 1999}}  \\ \hline\hline

\end{tabular}
\end{center}
\end{table}

\section{The High Resolution X-Ray Spectrum}

A portion of the dispersed high 
resolution X-ray spectrum of 
E0102-72 (a thousand year old supernova remnant in the 
Small Magellanic Cloud) is shown in Figure~1.
The Chandra observation was taken using the High Energy
Transmission Grating Spectrometer (HETGS) which employs the
High Energy Gratings in conjunction with the Advanced 
CCD Imaging Spectrometer (ACIS). The HETGS  separates the 
X-rays into their distinct X-ray emission lines, forming separate 
images of the remnant with the X-ray light of each line. 
The dispersed spectrum shows X-ray lines of highly ionized oxygen, neon,
magnesium and silicon. Several of the dispersed `ring' images differ 
from the zeroth order image. There are differences in radii, 
likely caused by the changing ionization state, that suggest
 the passage of the 
supernova ejecta through the reverse shock. Note also the distorted
shape of the NeX Ly~$\alpha$ line relative to the zeroth order. This
is due to Doppler shifts associated with high velocity material.
Both of these features are discussed in detail below.

\begin{figure}
\plotone{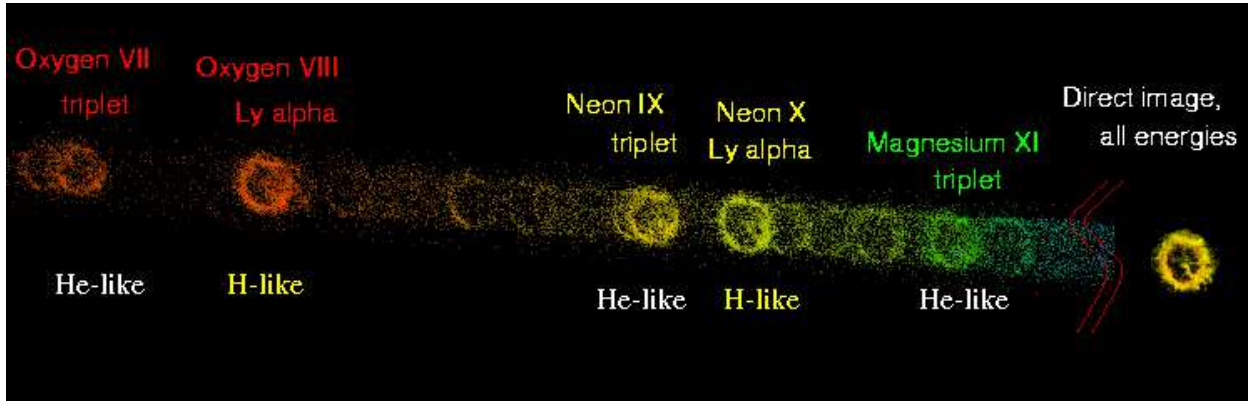}
\caption{Dispersed high resolution X-ray spectrum of E0102-72.
Shown here is a portion of the -1 order formed by the medium energy gratings,
color coded to suggest the ACIS energy resolution.  At right in the figure
is the zeroth order, which combines all energies in an undispersed image.
Differences in radii (as seen contrasting OVIII with OVII) and distortions 
in the shape (as seen in NeX Ly~$\alpha$) are significant, reflecting
progress of the reverse shock and Doppler shifts, respectively. The 
resonance line of helium-like silicon is not shown in the figure.}
\end{figure}

\section{The Ionization/Shock Structure of E0102-72: Measurements}

A side-by-side comparison of the dispersed OVII resonance 
line  with the OVIII Ly~$\alpha$ line is given in Figure~2.
OVII is helium-like oxygen, with two electrons remaining, 
whereas OVIII is hydrogen-like. The ring diameter of the OVIII 
line is obviously larger than that of the OVII line. 

\begin{figure}
\plottwo{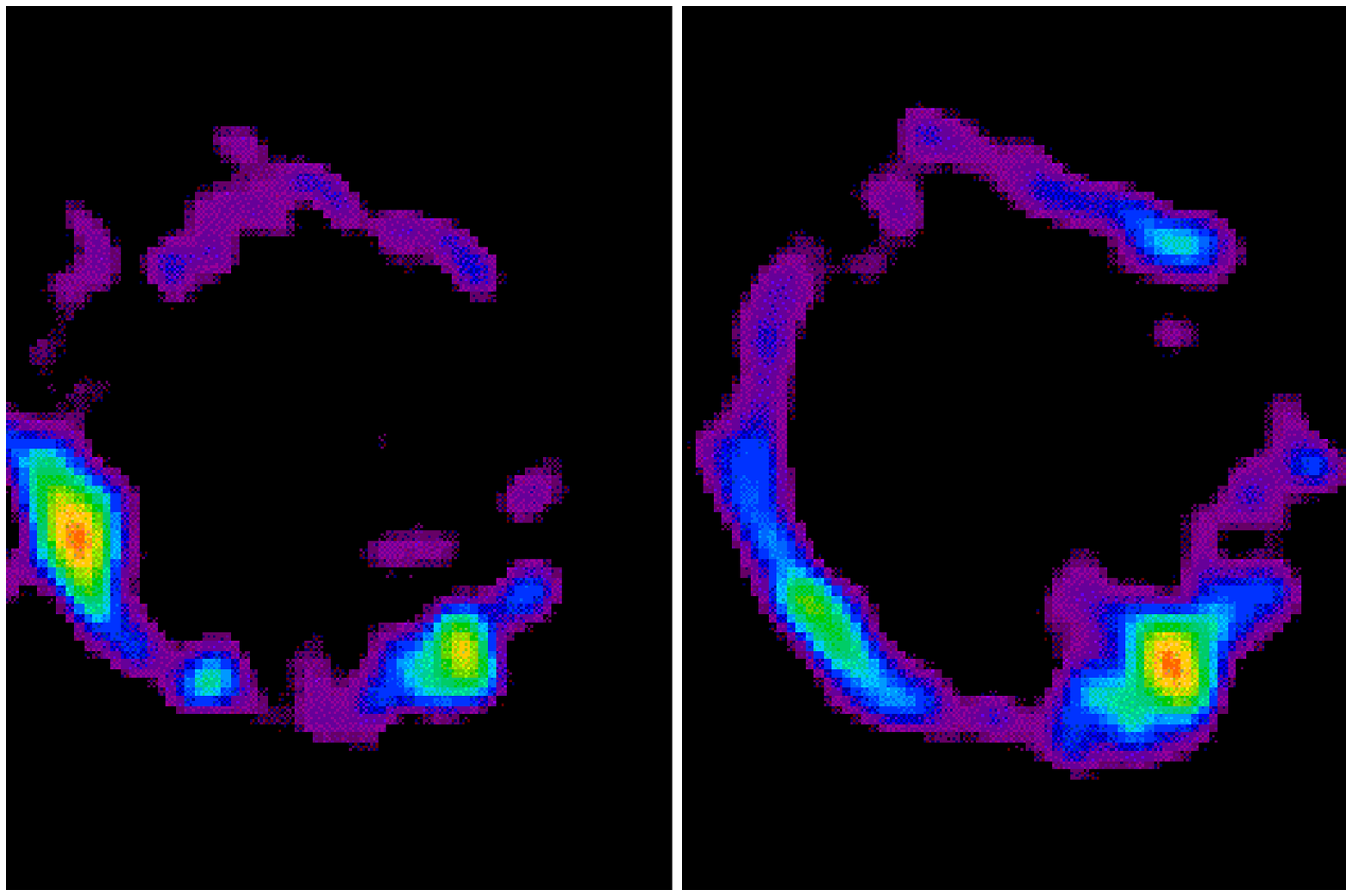}{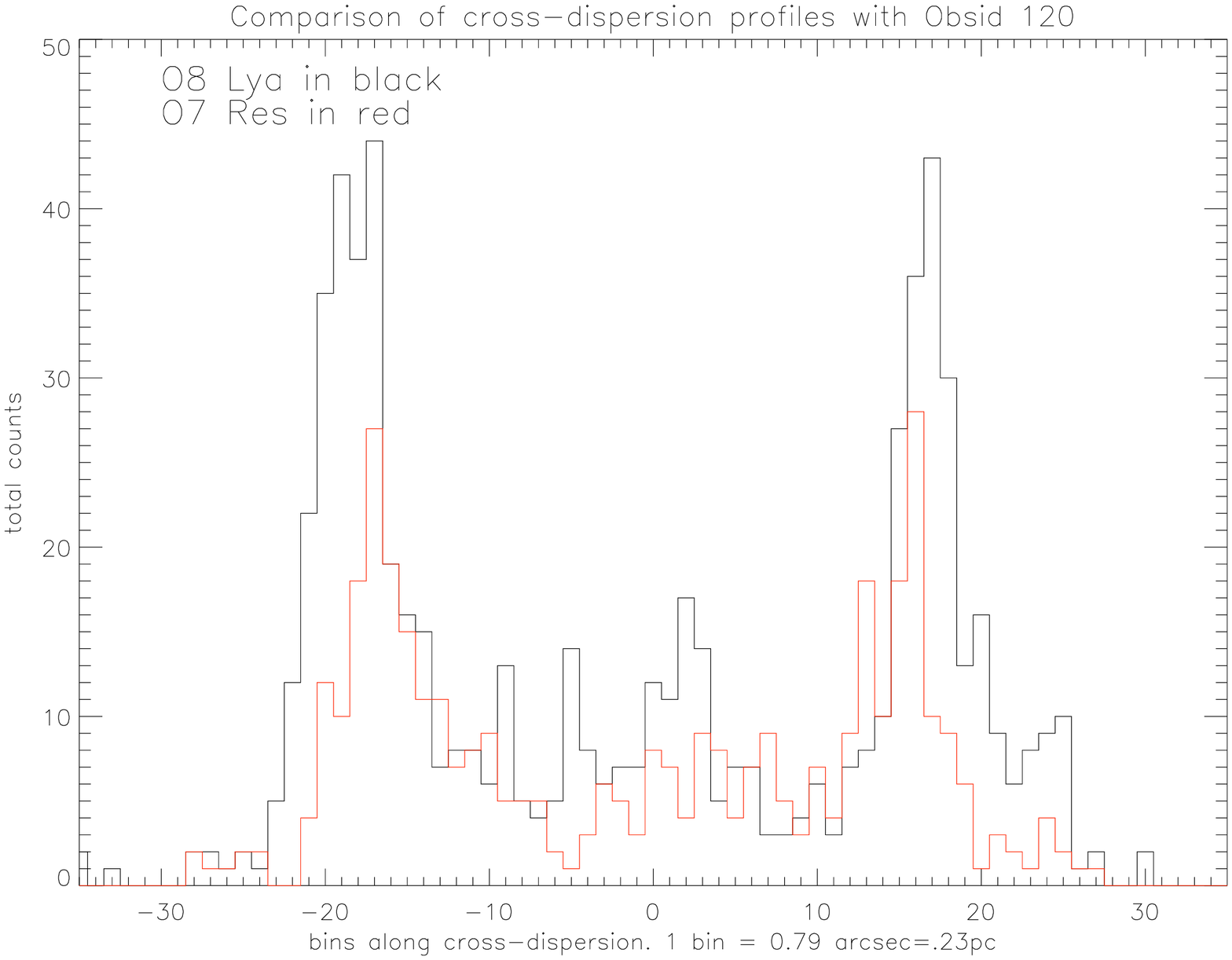}
\caption{The dispersed image formed by the OVII resonance line (574~eV) is
shown at far left next to that of the OVIII Ly~$\alpha$ line (654~eV), which
is shown to its immediate right on the same scale. 
The radius of the helium-like OVII line is obviously smaller than 
that of the hydrogen-like OVIII line. The right-hand panel overlays the
cross-dispersion histograms of these two images. The histogram  of
the OVII line (in red) is nestled inside that of the OVIII line (in black).
These histograms provided numerical measures of ring diameters as plotted
in Figure~3.}
\end{figure} 

The ring diameters for all the bright X-ray lines were measured by
tracing the distribution in the cross-dispersion direction, as illustrated
in the right panel of Figure~2, which shows the cross-dispersion histograms
for the OVIII Ly~$\alpha$ and OVII resonance lines. 
Each histogram begins below the bottom edge
of the ring and ends above the top edge. Clearly, the diameters of
the two rings differ markedly, with the helium-like OVII being narrower.

\begin{figure}
\plottwo{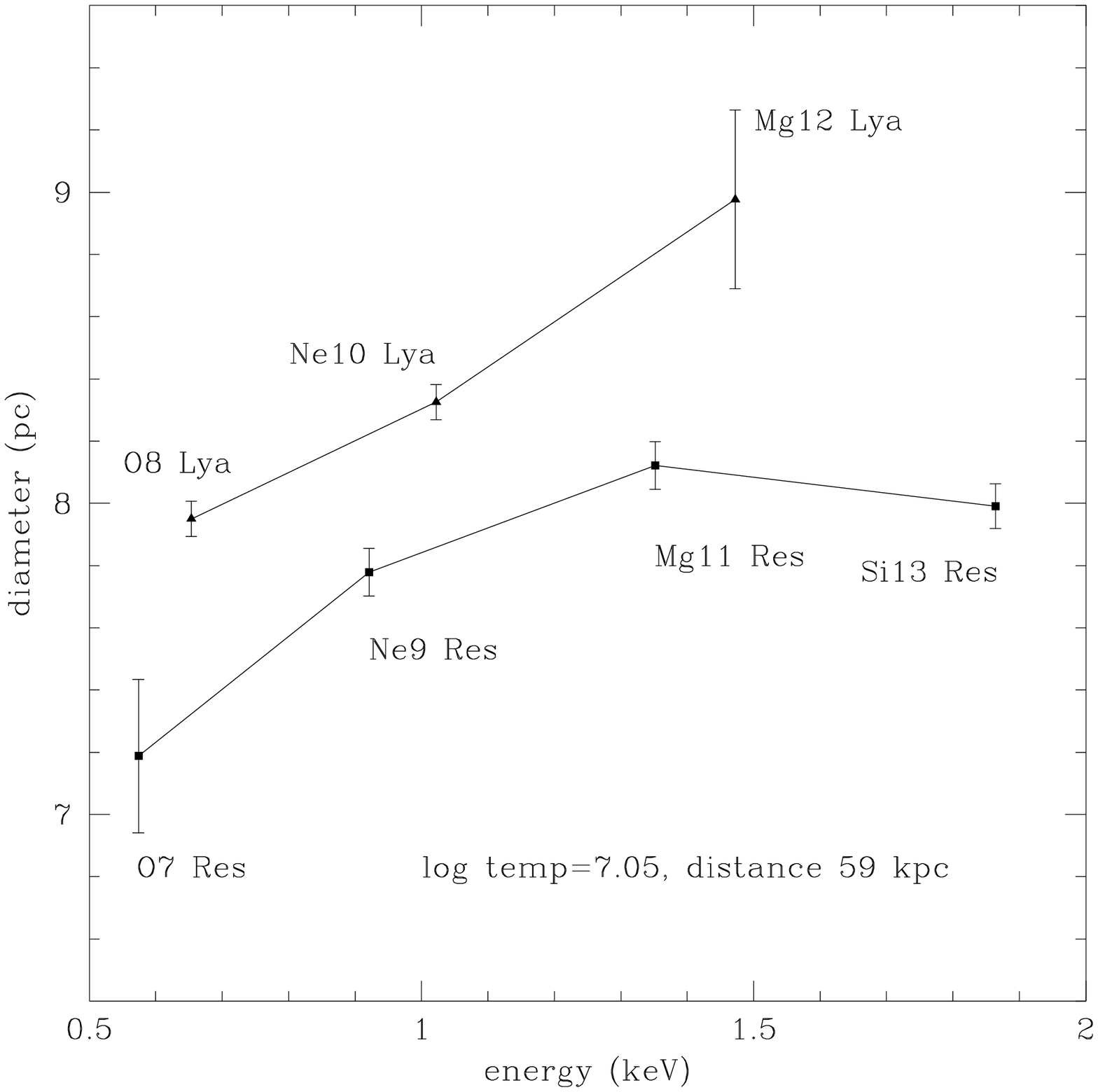}{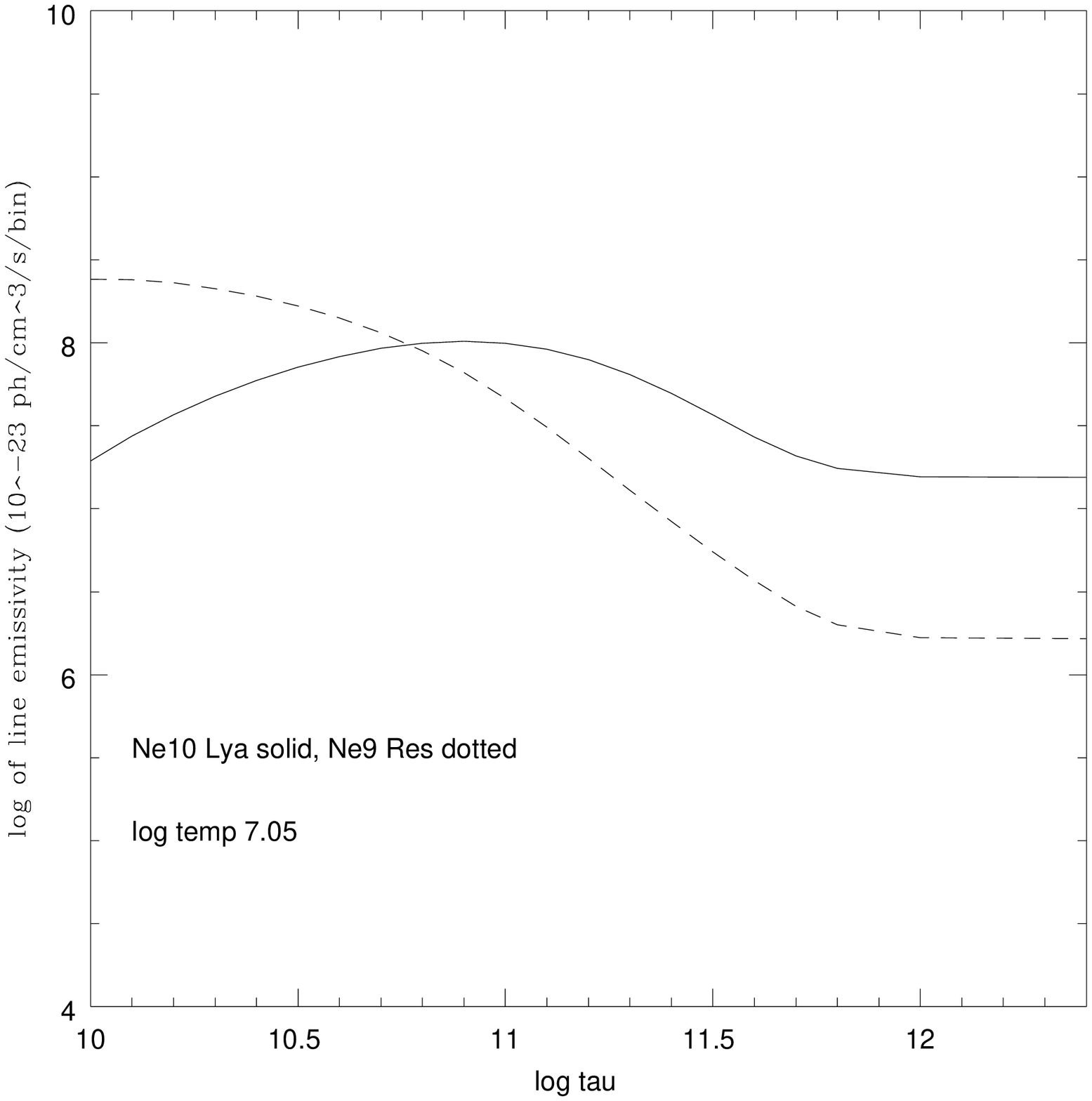}
\caption{{\bf Left:} The ring diameters for the bright lines are plotted 
against energy.
The top curve connects the hydrogen-like lines, and the bottom curve connects
the helium-like lines. Note that the hydrogen-like lines lie outward of
their helium-like counterparts, regardless of element. {\bf Right:} The 
emissivity of a helium-like line peaks earlier than that of a 
hydrogen-like line in an ionizing plasma.} 
\end{figure}

The measured diameters for the bright lines are shown in the Figure~3.
It is apparent that {\bf all of the hydrogen-like lines} (connected by the top 
curve) {\bf lie outside their corresponding helium-like lines} 
(connected by the bottom curve). As each element is considered on an
individual basis, the hydrogen-like emitting region lies outward of
the helium-like emitting region, suggesting a mechanism which is 
independent of spatial separation of the elements. This mechanism
is discussed in the next section.

\section{The Ionization/Shock Structure of E0102-72: Model}

Consider the evolution of an ionizing plasma toward equilibrium.
After the passage of a shock, at a fixed electron temperature 
$T_e$ and electron density $n_e$, a helium-like ionization state is 
created earlier than a hydrogen-like state. This is illustrated in the 
right panel of Figure~3, where a neon plasma has been heated to electron 
temperature $T_e$=10$^{7.05}$ by the passage of a shock. 
The emissivity of a line of helium-like neon is seen to peak
at an earlier ionization age than hydrogen-like neon 
(where the ionization age $\tau$ is defined as the product of electron 
density ne$_e$ and time since passage of the shock, i.e., $\tau$=n$_e$t).
The hydrogen-like outer region is therefore a more `evolved' plasma 
than the helium-like inner region. This suggests the action of a shock 
moving {\it inward} relative to the ejecta - the `reverse shock'- which 
is the standard model for the mechanism which heats SNR ejecta 
to X-ray temperatures. This is illustrated in the left 
panel of Figure~4.

The results of applying a very simple model are shown in the right
panel of Figure~4. The model assumes that the elements are
uniformly mixed, and that there is a single electron temperature. We
make these assumptions to examine the proposition that the ring 
diameter corresponds to ionization age, where we have estimated 
that age by finding the peak emissivity (Raymond \& Smith 1977;
Hughes \& Helfand 1985) for each bright X-ray line 
assuming a fixed T$_e$ of about 1~keV or log T$_e$=10$^{7.05}$ 
(as suggested by our global NEI analysis). 
Although any workable model must consider many 
parameters, the monotonic behavior of the data strongly suggests 
that these arcsec differences
in ring diameter are attributable to the ionization structure resulting
from the reverse shock.

\begin{figure}
\plottwo{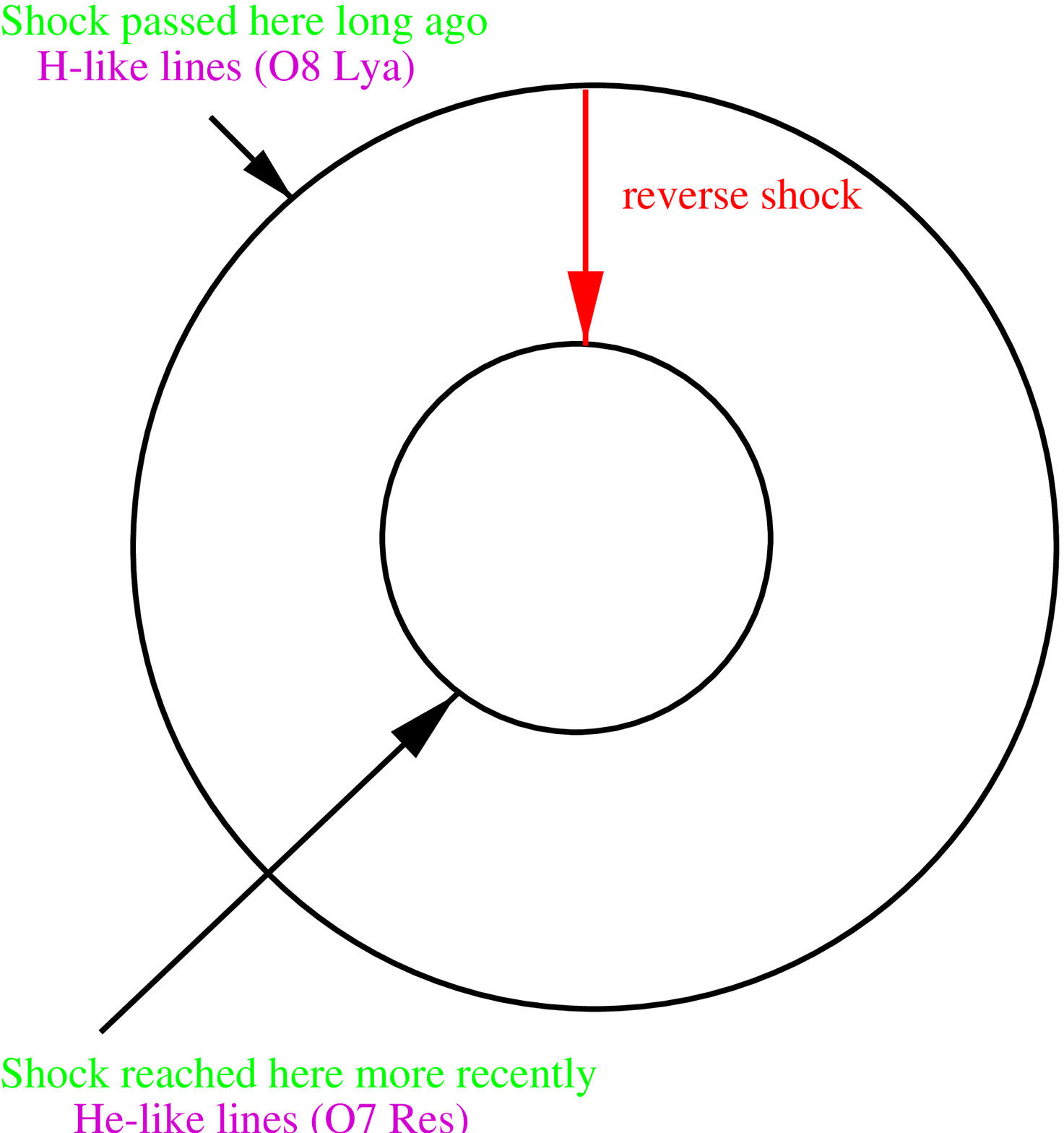}{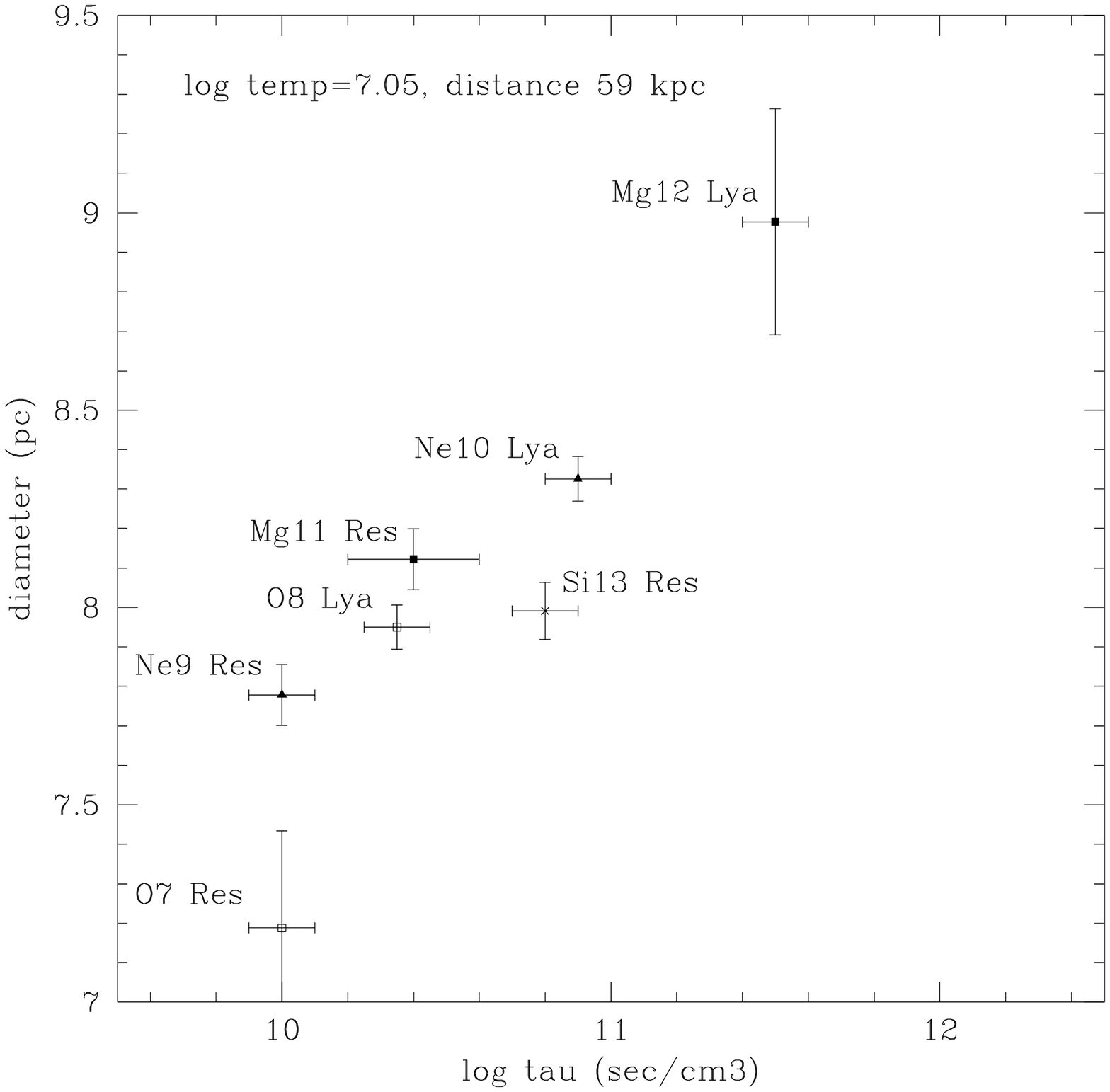}
\caption{The panel at left illustrates how the progress of the `reverse shock'
results in the more-evolved (hydrogen-like) plasma lying outward of
the less-evolved (helium-like) plasma. The panel at right shows
the results of applying of a simple model. Ring diameter and ionization
age are seen to be correlated, lending credence to the interpretation
that diameter differences are the result of ionization structure
resulting from the `reverse shock'. (The plotted point for OVII Res is 
an upper limit.)}
\end{figure}

\begin{figure}
\plotone{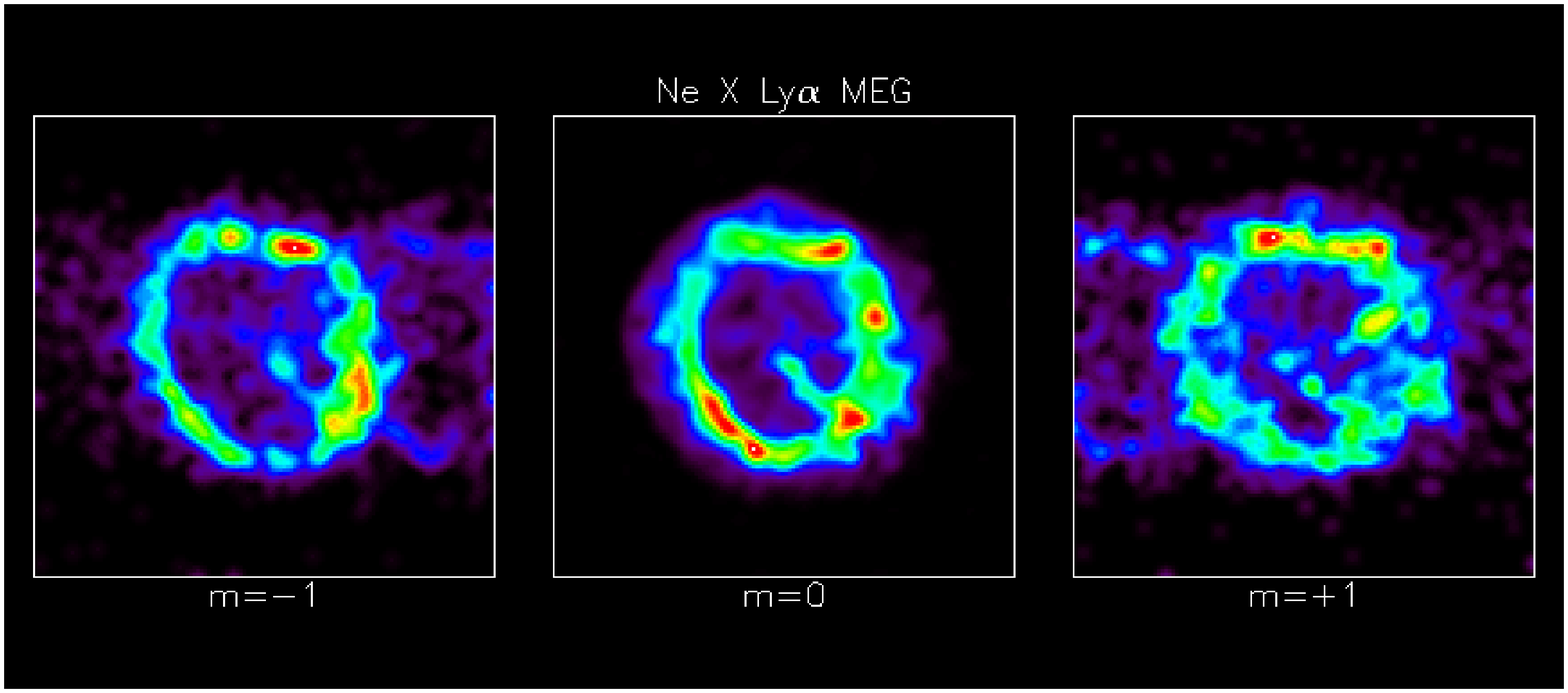}
\caption{The dispersed m=-1 and m=+1 orders of NeX~Ly~$\alpha$ 
show distortions relative to the undispersed zero (m=0) order. 
These distortions are attributed to Doppler shifts. }

\begin{center}
\begin{minipage}[h]{2.5in}
\plotone{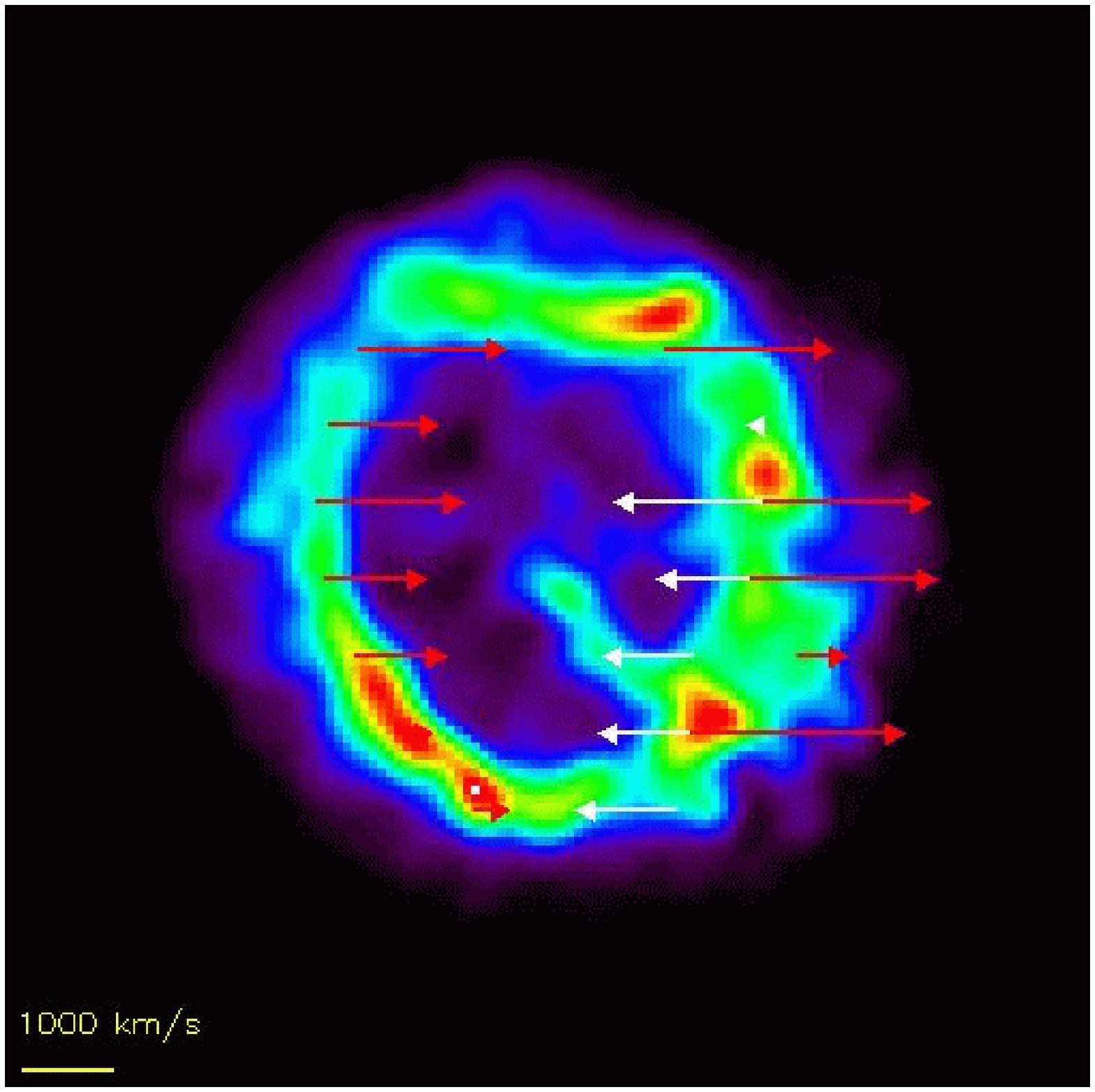}
\end{minipage}
\caption{Schematic Doppler map of E0102-72 as detected with NeX Ly~$\alpha$. 
The lengths and locations of the arrows are used to indicate 
the relative velocities and approximate locations associated with 
the Doppler shifts. White arrows pointing left represent blue shifts; red
arrows pointing right represent red shifts.}

\end{center}
\end{figure}

\section{The Velocity Structure of E0102-72}

	As seen in Figure~5, distortion is 
evident when dispersed images of NeX~Ly~$\alpha$ are compared with 
the zeroth order (m=0). (The zeroth order has been filtered to include 
only events around the 1022~eV energy
of NeX~Ly~$\alpha$). The leftmost panel shows the m=-1 ring, the middle
panel is the zeroth order, and the right panel is the m=+1 ring. Doppler
red shifts to longer wavelengths are indicated by distortions outward, away
from zero order, and blue shifts are indicated by distortions of the
ring {\it toward} the zeroth order at the center. Thus, the upper left sector
of the remnant appears to have a red shift as seen by examining the
m=-1 order ring, whereas the right half of the ring is broadened 
in the +1 order, indicating {\it both} red and blue shifts for the right
half of the remnant. KS tests support this qualitative 
interpretation, indicating that, to high confidence, the 
dispersion-direction profiles of the dispersed and undispersed 
images are {\it different}, while the cross-dispersion profiles are 
{\it the same}.

	The results of quantitative analysis are schematically 
illustrated in Figure~6, where white arrows 
pointing toward the left indicate blue shifts, and red arrows pointing 
toward the right indicate red shifts. The lengths and locations
of the arrows are used to indicate the relative velocities and 
approximate locations associated with the Doppler shifts.
Typical red and blue-shifts indicate velocities on the order of 1000~km/s 
(Houck et al. 2000), comparable to those of optical filaments.

\section{Conclusions}

The {\it Chandra} high-resolution X-ray spectrum of E0102-72 obtained
with the HETGS reveals images of the supernova remnant in the light
of individual lines of O, Ne, Mg and Si. Radial variations among different
ionization stages are seen on an arcsecond spatial scale, consistent
with an evolving ionization structure due to the passage of
the ejecta through the reverse shock. Doppler shifts of
$\sim$1000~km/sec (comparable to the velocities of optical 
filaments) have been detected at various locations along the remnant. 
This velocity information is being applied toward a model
of the morphology and motion of the X-ray emitting material.

\acknowledgements
We thank Glenn Allen, Norbert Schulz, Tom Pannuti and Sara-Anne Taylor
for helpful discussions. We are grateful to the CXC group at MIT
for their assistance in analysis of the data. This work was 
prepared under NASA contract NAS8-38249 and SAO SV1-61010.

\end{document}